**Numerical simulation of a comparative study on heat extraction from Soultz-sous-Forêts geothermal field using supercritical carbon dioxide and water as a working fluid**

Mrityunjay Singh [1, *], Saeed Mahmoodpour [2, *], Reza Ershadnia[3], Mohamad Reza Soltanian [3], Ingo Sass[1,4]

[1] Group of Geothermal Science and Technology, Institute of Applied Geosciences, Technische Universität Darmstadt;
[2] Group of Geothermal Technologies, Technical University of Munich;
[3] Department of Geology, University of Cincinnati, Cincinnati, USA;
[4] Helmholtz Centre Potsdam, GFZ German Research Centre for Geosciences;

**\*Correspondence:** Mrityunjay: mrityunjay.singh@tu-darmstadt.de; Saeed: saeed.mahmoodpour@tum.de

**Abstract:** Geothermal energy is an infinite energy source for the present human society. Energy extraction from the deep subsurface requires engineering using a working fluid that circulates between well doublet. Due to its thermal properties, $CO_2$ is an ideal option as a heat transfer fluid. By using $CO_2$, working fluid loss is an advantage compared to other working fluids. This study developed a field-scale hydro-thermal model to examine the heat extraction potential from Soultz-sous-Forêts with $CO_2$ as the working fluid. Results are compared for the same scenario with water as the working fluid. A better understanding of the heat extraction mechanism is established by considering the reservoir response and the wellbore heat exchange. Sensitivity analyses are performed for different injection temperatures and flow rates for 50 years. Results show that the wellbore effect is multiple times higher than the reservoir response to the production temperature. Furthermore, lowering the injection temperature eventuates to a smaller temperature reduction at the subsurface, enhancing the overall heat extraction potential with a minor impact on thermal breakthrough. The cold region developed around the injection wellbore may affect the production fluid temperature due to its proximity to the production wellbore. To reach higher heat extraction efficiency, it is essential to use sufficient wellbore spacing. $CO_2$ can be used as working fluid for over 50 years as it does not show significant thermal breakthrough and temperature plume evolution in the reservoir under studied conditions. $CO_2$ shows lower temperature reduction for all injection rates and temperatures for 50 years of operation.

**Introduction**

$CO_2$ emission into the atmosphere has catastrophic impact on climate change. Recent studies by IPCC indicates an increase in 2.7 ℃ temperature over pre-industrial temperature level by the end of this century [IPCC, 2022]. To mitigate this impact, renewable energy resource is promoted. On the other hand, $CO_2$ sequestration in the subsurface formations, has a potential to reduce its emission from existing fossil fuel power plants [Mahmoodpour et al. 2019a,b; Singh et al. 2020a, 2021; Hou et al. 2022; Wang et al. 2022]. Possibility of using $CO_2$ as working fluid for geothermal energy extraction has two advantages. (a) Studies regarding thermal properties of $CO_2$ shows it has lesser viscosity, and smaller heat capacity than water for certain range of temperature and pressure. $CO_2$ has much smaller geochemical reactivity compared to water. At the same time, expansivity of $CO_2$ reduces the fluid pumping cost compared to water [Pruess 2006; Singh et al. 2020b]. (b) $CO_2$ loss has the advantage for anthropogenic $CO_2$ geosequestration. This makes the $CO_2$ a suitable candidate for heat transmission fluid from geothermal reservoirs. $CO_2$ as a geothermal heat carrier is widely studied regarding two major geothermal exnery extraction processes: $CO_2$ plume geothermal (CPG) [Adams et al. 2014; Buscheck et al. 2016; Ezekiel, J., 2021, 2022] and $CO_2$ Enhanced Geothermal System [Liu et al. 2003; Pruess 2006, 2008; Luo et al. 2013, 2014]. In this paper, a hydro-thermal numerical model is developed for demonstrating the feasibility of using supercritical $CO_2$ as working fluid for one of the working geothermal field, Soultz-sous-Forêts, France. Furthermore, energy extraction potential is compared with water for the same geological setting.

Brown [2000] proposed the concept of using supercritical $CO_2$ as the working fluid for a geothermal system. This study suggests that use of $CO_2$ results in small geochemical reactivity and consequently lower scaling and corrosion problems in the wellbore tubing and pipes. The expansivity of $CO_2$ is a function of temperature and pressure, and resulting density difference between the injection and production wells reduces the pumping load. Furthermore, there is a potential to use $CO_2$ in supercritical geothermal systems. Detailed chemical reactions with

$CO_2$ and geothermal rock systems reveal that the dissolution enhances with $CO_2$ content [Liu et al. 2003]. Comparing $CO_2$ with water showed that heat extraction rate and subsurface temperature increases [Pruess, 2006]. To avoid thermal breakthrough Pruess [2008] proposed a well placement in a way that the production well is located at the top section of the reservoir. This process is more effective when $CO_2$ is used as working fluid compared to water. Pressure dependent mobility is examined by Zhang et al. [2014] and they concluded that it is a favorable parameter for $CO_2$ in comparison to water. $CO_2$ has lower specific heat capacity in comparison to water and it requires higher mass flux to reach the same heat extraction rate in comparison to water, however the pressure gradient between the wells would be smaller for $CO_2$ [Luo et al. 2014]. Later studies demonstrated that $CO_2$ reaches to a higher production temperature for the same mass flux [Liu et al. 2017]. Numerical simulations for the fractured reservoirs show that increasing the mass flux and fracture roughness eventuates to a higher heat transfer efficiency for $CO_2$ system [Bai et al. 2018]. For a two-phase system, increasing the $CO_2$ content eventuates to a better production capability [Chen et al. 2019].

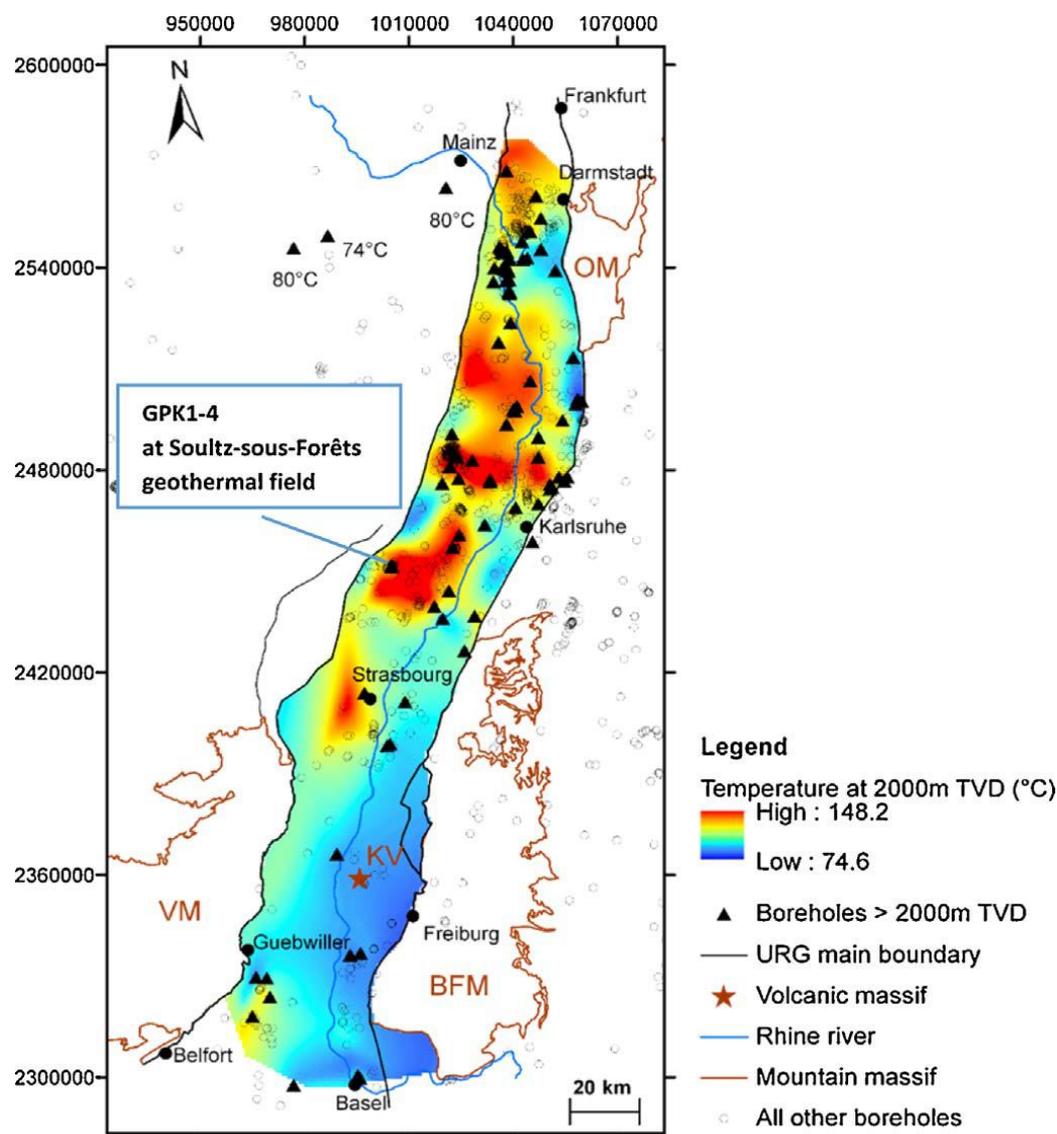

*Figure 1: Soultz-sous-Forêts and its surrounding region temperature distribution at 2 km depth TVD in the Upper Rhine Graben [Baillieux et al. 2013].*

Soultz-sous-Forêts is a geothermal site which is located in France. The geothermal gradient for this reservoir is abnormally high for the depth between 1 and 3 km. Figure 1 shows temperature distribution in the vicinity of the wellbores (GPK1, GPK2, GPK3 and GPK4) which shows high temperature zones in the vicinity of Soultz-sous-Forêts geothermal site [Baillieux et al. 2013]. GPK1 is not considered in this study in accordance with recent field operations [Mahmoodpour et al. 2021]. From the geological point of view, the top section is created by thick

Quaternary and Tertiary sediments, whereas middle section is made of Mesozoic to Paleozoic sedimentary rocks. The bottom section is made of crystalline granite [Duringer et al. 2019]. The bottom section which is the point of interest in this study, located between 1420 to 4700 m. This region comprises monzogranite with K-feldspar mega crystal with localized concentration of biotite. For the further deeper region two-mica granite is available between 4700 and 5000 m [Cocherie et al. 2004; Schill et al. 2018]. Measurements of the rock thermophysical and petrophysical properties shows that the hydraulic conductivity is $5\times10^{-8}$, $1\times10^{-8}$ and $9\times10^{-9}$ m/s for upper, mid and bottom sections, respectively. Correspondingly, the specific storage for these layers are $8\times10^{-7}$, $5\times10^{-7}$, and $1.75\times10^{-8}$ $m^{-1}$. The porosity and thermal conductivity of the upper sediment are 0.1 and 2.8 W/m/k whereas the mid layer sediments and the crystalline granite have same porosity (0.03) and thermal conductivity (2.5 W/m/k). The thermal capacity of the top layer is lowest ($2\times10^6$ $J/m^3/K$) followed by the lower layer ($2.9\times10^6$ $J/m^3/K$) and the mid layer has the highest thermal capacity ($3.2\times10^6$ $J/m^3/K$) [Rolin et al. 2018].

Studies for Soultz-sous-Forêts reservoir characterization resulted to the 39 fracture zones [Sausse et al. 2010; Dezayes et al. 2010]. The fracture zones (faults) demonstrating the intersection with a wellbore have more contribution on the fluid flow and heat transfer. From these 39 fracture zones, only five have intersection with at least one of the wellbores and therefore, these 5 faults are considered for the numerical simulations [Rolin et al. 2018]. Details of these faults are listed in Table 1 and the corresponding geometrical arrangement is depicted in Figure 2.

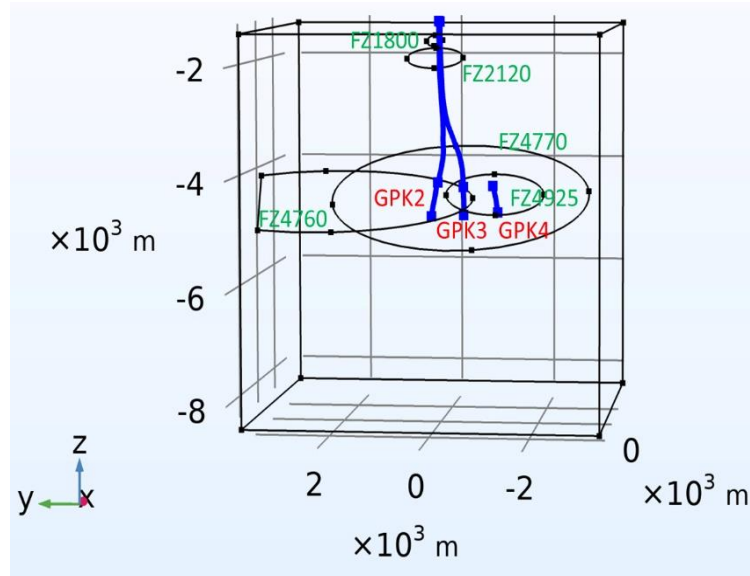

*Figure 2: Geometry of the Soultz-sous-Forêts site considered for numerical simulations. The FZ stands for fault zone, GPK3 and GPK4 are the injection wells, GPK2 is the production well.*

*Table 1: Fault parameters at in-situ conditions [Rolin et al. 2018; Mahmoodpour et al. 2021].*

| Parameter | Unit | FZ1800 | FZ2120 | FZ4760 | FZ4770 | FZ4925 |
|---|---|---|---|---|---|---|
| Hydraulic conductivity | m·s$^{-1}$ | $6.08\times10^{-6}$ | $1.7\times10^{-5}$ | 0.05 | $2\times10^{-5}$ | $6.3\times10^{-5}$ |
| Specific storage | 1·m$^{-1}$ | $2\times10^{-6}$ | $2\times10^{-6}$ | $2\times10^{-6}$ | $2\times10^{-6}$ | $2\times10^{-6}$ |
| Porosity | - | 0.1 | 0.1 | 0.1 | 0.1 | 0.1 |
| Thermal conductivity | W·m$^{-1}$·K$^{-1}$ | 2.5 | 2.5 | 2.5 | 2.5 | 2.5 |
| Thermal capacity | J·m$^{-3}$ K$^{-1}$ | $2.9\times10^{-6}$ | $2.9\times10^{-6}$ | $2.9\times10^{-6}$ | $2.9\times10^{-6}$ | $2.9\times10^{-6}$ |
| Thickness | m | 12 | 15 | 8 | 15 | 1 |

In this simulation study, GPK2 is used as the production well and GPK3 & GPK4 (see Figure 2) are the injection wells which is the same operational arrangement considered for heat extraction using water between June 2016 and September 2019 [Mahmoodpour et al. 2021]. These wells are aligned alongside the NNW-SSE direction which coincides with the main fault direction. GPK2 and GPK3 shows leakage along the long section of the cased zone. Granitic reservoir section is open for all three wells. A wide range of studies are performed for static and dynamic modeling of Soultz-sous-Forêts geothermal reservoir. Analytical dispersive transfer model is used to characterize the flow circulation between GPK2 and GPK3 [Sanjuan et al. 2006]. Similar studies are performed to examine the connectivity between the wells using the tracer tests [Blumenthal et al. 2007; Gessner et al. 2009; Egert et al. 2020]. To recognize the hydraulic conductivity and the fracture sets, Gentier et al. [2010] used a particle tracking methodology for discrete fracture network of Soultz-sous-Forêts reservoir. Two-dimensional thermo-hydro-mechanical simulations were developed by Magnenet et al. [2014] and Vallier et al. [2020] at reservoir scale for water as the working fluid. However, availability of the 3D data enables us to examine the process in a more realistic condition with the original reservoir.

The heat exchange between the fluid and the wellbore is an important factor for the accurate estimation of heat extraction from a geothermal system. Heat exchange in the wells have detailed literature existence for enhanced oil recovery projects through the thermal operations [Hasan et al. 2002, 2009; Moradi et al. 2013, 2020]. Detailed knowledge regarding the temperature profile along the injection and production wells is a requirement for precise determination of injection bottom hole temperature and the surface production temperature. This is a missing point in most of the previous studies regarding heat extraction from the geothermal systems. High computational cost requirements for fully coupled wellbore effects and the reservoir response makes this problem uneconomical. Possibility of coupling of the analytical model for the wellbore response with the numerical simulations of the reservoir seems to be a solution for this problem.

In this study, a thermo-hydraulic model is developed with $CO_2$ as the working fluid based of real field data [Mahmoodpour et al. 2021]. A detailed analysis of discrete fault structures and its impact on the fluid dynamic behavior is critically analyzed here. Heat exchange of the fluid inside the injection, production wells and the surrounding formations is combined with the reservoir thermal response. Further, sensitivity analysis is performed with different injection temperature and mass flow rates to examine the efficiency of the colder fluid injection and the heat extraction potential at Soultz-sous-Forêts. To consider the heat exchange between the supercritical $CO_2$ and the wellbore, the analytical model developed by Hasan et al. [2009] is combined with the reservoir scale 3D hydro-thermal simulations based on available field data.

**Methodology**
The computational method used for modeling the Soultz-sous-Forêts reservoir is discussed below. Reservoir modeling involving supercritical $CO_2$ and water transmission through the faults and the rock matrix are discussed first followed by an analytical model for the wellbore heat exchange.

**Hydro-thermal modeling**
The mass balance equation coupled with heat transfer in a fractured reservoir is modeled using the following equation (COMSOL):

$$\rho_1(\phi_m S_1 + (1-\phi_m)S_m)\frac{\partial p}{\partial t} - \rho_1\big(\alpha_m(\phi_m \beta_1 + (1-\phi_m)\beta_m)\big)\frac{\partial T}{\partial t} = \nabla.\left(\frac{\rho_1 k_m}{\mu}\nabla p\right) \quad [1]$$

Fluid pressure and temperature as the independent variables are implemented in this modeling approach and they designated by $p$ and $T$ respectively. Storage effect is considered by taking storage coefficient of $S_1$ and $S_m$ for the fluid and rock, respectively. In the above equation, $\alpha_m$ is the Biot's coefficient of porous media. It is assumed that the porosity of rock remains constant at $\phi_m$ during the entire simulation. Since the porosity is constant for the entire simulation period, a constant value of the reservoir permeability ($k_m$) is implemented. Volume of the fluid and the rock may change by the temperature through the $\beta_1$ and $\beta_m$, as the thermal expansion coefficients. Thermophysical properties including density ($\rho_1$), viscosity ($\mu$), thermal conductivity ($\kappa$), specific heat capacities ($C_p$ & $C_v$) are function of pressure and temperature (Span and Wagner, 1996). For the thermophysical properties of water, following equations are used (COMSOL):

$$\mu = 1.38 - 2.12 \times 10^{-2} \times T^1 + 1.36 \times 10^{-4} \times T^2 - 4.65 \times 10^{-7} \times T^3 + 8.90 \times 10^{-10} \times T^4 - 9.08 \times 10^{-13} \times T^5 + 3.85 \times 10^{-16} \times T^6 (273.15 - 413.15K) \quad [2.1]$$
$$\mu = 4.01 \times 10^{-3} - 2.11 \times 10^{-5} \times T^1 + 3.86 \times 10^{-8} \times T^2 - 2.40 \times 10^{-11} \times T^3$$

$$(413.15 - 553.15K) \qquad [2.2]$$

$$C_p = 1.20 \times 10^4 - 8.04 \times 10^1 \times T^1 + 3.10 \times 10^{-1} \times T^2 - 5.38 \times 10^{-4} \times T^3 + 3.63 \times 10^{-7} \times T^4 \qquad [3]$$

$$\rho = 1.03 \times 10^{-5} \times T^3 - 1.34 \times 10^{-2} \times T^2 + 4.97 \times T + 4.32 \times 10^2 \qquad [4]$$

$$\kappa = -8.69 \times 10^{-1} + 8.95 \times 10^{-3} \times T^1 - 1.58 \times 10^{-5} \times T^2 + 7.98 \times 10^{-9} \times T^3 \qquad [5]$$

Heat flux at the bottom of the boundary domain is implemented that is similar to Mahmoodpour et al. [2021] and its magnitude is 0.07 W/m². The open boundaries for heat and mass to the side walls are considered. A small part of the reservoir is considered for modeling. Therefore, fault FZ4760 with its high conductivity contribute mainly to heat and mass transfer into/from the outer boundaries as it cuts the side boundaries. Large scale faults as shown in Figure 2 are considered as a homogeneous medium with the mentioned porosity and hydraulic conductivity in Table 1. During the simulation, they are treated like internal boundaries which is common for the fracture effect implementation to the heat and mass flux coupling (in the remaining section, the fractures instead of the faults for the formulation and simulations are assumed). The heat and mass exchange between the fracture and the matrix zone is considered in the following equation:

$$\rho_l \big( \phi_f S_1 + (1-\phi_f) S_{mf} \big) e_h \frac{\partial p}{\partial t} - \rho_l \Big( \alpha_f \big( \phi_f \beta_1 + (1-\phi_f) \beta_f \big) \Big) e_h \frac{\partial T}{\partial t} = \nabla_T . \left( \frac{e_h \rho_1 k_f}{\mu} \nabla_T p \right) + n. Q_m \qquad [6]$$

The subscript $f$ is used to make difference between the fracture and matrix, and the nomenclature are the same as Eq. [1]. Hydraulic aperture of the fracture which is the most important factor for the heat extraction from the geothermal system [Mahmoodpour et al. 2022a, b] is shown by $e_h$. Fracture permeability ($k_f$) is calculated based on the hydraulic conductivity which as listed in Table 1. Fracture length is much large than the fracture width and therefore, changes in the mass and heat are tracked alongside the fracture lengths and $\nabla_T$ shows the gradient operator. The possible mass flux between fracture and the matrix zone is inserted with $n. Q_m = n. \left( -\frac{\rho k_m}{\mu \nabla p} \right)$ through the simulation and in this way, coupling between the fracture and matrix is considered. Well diameter is negligible in comparison to the overall size of the system. To make the simulations computationally feasible and avoiding the very fine mesh grid for the wellbore, wellbore is assumed as a line source for the heat and mass transmission in the simulation.

Since there is a time lag between the temperature development inside the fracture and matrix zone in the discrete fracture model, therefore local thermal non-equilibrium methodology is adopted here to track the heat transfer between $CO_2$ (or water) and the rock. Considering a local thermal equilibrium approach would accelerate the computational simulation however, there would be an overestimation in the production rate [Shaik et al. 2011]. Similar to the mass transfer, two equations are used to model a heat transfer inside the matrix zone and the fractures. Coupling between them is implemented through the exchange term. For the matrix zone:

$$(1-\phi_m)\rho_m C_{p,m} \frac{\partial T_m}{\partial t} = \nabla . \big( (1-\phi_m) \lambda_m \nabla T_m \big) + q_{ml}(T_l - T_m) \qquad [7]$$

Based on the local thermal non-equilibrium methodology, $T_m$ and $T_l$ as the temperature for the matrix and supercritical $CO_2$ (or water) are considered. Thermal properties and the rock density are less sensitive on the temperature in comparison to the $CO_2$ (or water), therefore constant values of $\rho_m$, $C_{p,m}$, $\lambda_m$ and $q_{ml}$ are used for rock density, rock specific heat capacity, rock thermal conductivity and the rock- $CO_2$ (or water) heat transfer coefficient, respectively. Again, the subscript $f$ is used to make a difference between the matrix and the fractures in the heat transfer equation as shown below:

$$(1-\phi_f) e_h \rho_f C_{p,f} \frac{\partial T_m}{\partial t} = \nabla_T . \big( (1-\phi_f) e_h \lambda_f \nabla_T T_m \big) + e_h q_{fl}(T_l - T_m) + n. \big( -(1-\phi_m) \lambda_m \nabla T_m \big) \qquad [8]$$

Eq. [9] is used to obtain the $CO_2$ (or water) temperature inside the matrix zone.

$$\phi_m \rho_l C_{p,l} \frac{\partial T_l}{\partial t} + \phi_m \rho_l C_{p,l} \left(-\frac{k_m \nabla p}{\mu}\right) . \nabla T_l = \nabla . (\phi_m \lambda_l \nabla T_l) + q_{ml}(T_m - T_l) \qquad [9]$$

As it is mentioned, $CO_2$ (or water) specific heat capacity ($C_{p,l}$) and thermal conductivity are updated based on temperature before the next step calculation. Similar methodology is used to track the $CO_2$ (or water) temperature inside the fracture zone but the heat flux as $n.q_l = n.(-\phi_l \lambda_l \nabla T_l)$ is used to insert the heat exchange between the $CO_2$ (or water) and the fractures.

$$\phi_f e_h \rho_l C_{p,l} \frac{\partial T_l}{\partial t} + \phi_f e_h \rho_l C_{p,l} \left(-\frac{k_f \nabla_T p}{\mu}\right) . \nabla_T T_l = \nabla_T . (\phi_f e_h \lambda_l \nabla_T T_l) + e_h q_{fl}(T_m - T_l) + n.q_l \qquad [10]$$

COMSOL Multiphysics version 5.5 [COMSOL] is used for developing a reservoir scale thermo-hydraulic model for Soultz-sous-Forêts. Free tetrahedral meshes are considered for descretization of the geometry. The maximum element size in the matrix and the fault zone is 400 m, whereas the minimum element size is 0.7 m. The element growth rate cutoff is 1.3 to ensure faster convergence. For the region in the vicinity of the wellbores, maximum element size is considered as 0.7 m to ensure that the maximum mesh element size is smaller than the least fault zone thickness. The total number of domain elements are 5818288, boundary elements are 24909 and edge elements are 11675. Absolute tolerance value of $10^{-8}$ is used for all the simulations. Backward differentiation formula (BDF) is considered for numerical discretization and automatic time stepping is considered. Overall methodology is validated in previous studies for water as the working fluid by Mahmoodpour et al. [2021].

**Wellbore heat exchange with the formation**
When fluid move along the wellbore, there is heat exchange between the fluid and the formation. Based on this fact, the injected fluid reaches to a high temperature when it arrives to the bottom of the wellbore and producing fluid experiences heat loss when it reaches to the wellhead. Recent study [Mahmoodpour et al. 2021] show that this heat exchange is the controlling parameter on heat extraction rate from Soultz-sous-Forêts, by using water as the working fluid. In that study, availability of the operational data for water enabled the researchers to tune the required parameters for the wellbore effects. Here, in this study, similar approach is adopted but due to the lack of operational data as $CO_2$ being the working fluid, a sensitivity analysis is performed for the reasonable range of parameters to combine the effect of wellbore with the reservoir response. To do this, the wellbore flow model that is developed by Hasan et al. [2009] is used. This analytical model is based on the steady state condition. It enables us to consider different deviation for the wellbore. By considering $\alpha$ as the deviation angle between the wellbore and the horizontal axis in each section of wellbore, the following equation is used to calculate the fluid temperature inside the wellbore.

$$T_f = T_{ei} + \frac{1 - e^{(z-L)L_R}}{L_R} \left[g_G \sin\alpha + \Phi - \frac{g \sin\alpha}{c_p}\right] \qquad [11]$$

In the above equation $T_f$ is the fluid temperature at each depth of the system, $T_{ei}$ is the rock temperature at that depth, $z$ is depth, $L_R$ is the relaxation parameter, $g_G$ is the geothermal gradient, $\Phi$ is the lumped parameter for kinetic energy and Joule-Thomson coefficient (accounts for the casing properties, cement properties and their thicknesses). For derivation of this equation and more details of these parameters see Hassan et al. [2009] and Mahmoodpour et al. [2021].

So far, the fluid temperature at each depth inside the wellbore are available. Furthermore, mass flux at the same depth based on the fluid velocity and density can be calculated. Using the following equation, the output temperature can be calculated as:

$$T_{out} = \frac{\int m\, c_p\, T\, dz}{\int m\, c_p\, dz} \qquad [12]$$

This approach is used for the production temperature of GPK2 and injection temperatures of GPK3 and GPK4. Two value of $\Phi = 0.3$ K/m and 0.4 K/m are used to understand the sensitiveness of this lumped parameter over the final temperature whereas all the remaining simulations are performed considering 0.35 K/m. The fluid is single phase supercritical $CO_2$ flow and the model parameters are constant specific heat capacity of water as 1600

J/kg/k. For $CO_2$, $L_R$ is a function of mass flux and it show negligible impact on the production temperature. Therefore, a single value of $L_R$ is used for each flow rate.

**Results and Discussions**

Numerical models developed for geothermal energy extraction from Soultz-sous-Forêts using supercritical $CO_2$ and water as working fluids are presented in this section. This numerical methodology is validated against analytical models [34] and operational model of the Soultz-sous-Forêts [Mahmoodpour et al. 2021]. In this paper, reservoir modeling for $CO_2$ (or water) flow and heat transfer coupled with analytical model for wellbore heat exchange is shown. To examine the sensitivity of the mass flux, four levels of the mass flux as 20, 30, 40 and 50 kg/s from each injection well (GPK3 and GPK4) is assumed during the numerical simulation. Therefore, the produced mass flux from the GPK2 would be 40, 60, 80 and 100 kg/s for these cases respectively. Furthermore, to examine the possibility of reducing injected fluid temperature at the surface to maximize the heat extraction potential and by avoiding the thermal breakthrough time, four levels of injection temperature at the surface as 40, 50 60 and 70 °C are considered. The injection rates and the temperatures are in the similar range from the past operational values [Mahmoodpour et al. 2021]. 50 years is selected as the long-term operational period for this system. Temperature profile along the injection wellbores for different flow rates and injection temperatures are measured. As the interaction between $CO_2$ and wellbores for heat exchange is unknown due to unavailability of experimental and field data, sensitivity analysis is done for $\Phi = 0.3$ K/m and 0.4 K/m.

Figure 3 shows the temperature profile alongside the injection wells. Temperature profile of GPK3 involves two different sections: the shallower one with lower temperature gradient demonstrates $CO_2$ flow through the casing zone and the deeper one has open hole section along the wellbore. With the lower flow rate, at the casing zone, the heat exchange between the wellbore and the reservoir is small and therefore a lower temperature gradient is observed. For GPK4, $CO_2$ flow only through the open hole zone and the temperature profile alongside this wellbore indicates a quasilinear behavior. As it is obvious with reducing the flow rate, the available time for heat exchange between the fluid and the reservoir during this movement through the wellbore increases and it arrives with higher temperature to the bottom hole. For example, for GPK4, with injecting temperature of 70 °C, the bottom hole temperature with 20 kg/s would be approximately 4 °C higher than 50 kg/s. Similar comparison for GPK3 results to approximately 5 °C higher than 50 kg/s. Another observation regarding the wellbore effect is that the temperature difference between the maximum and minimum injection temperature at the surface reduces significantly when it is measured at the bottom. For example, in Figure 3(a1), the surface temperature difference is 30 °C (maximum is 70 °C and minimum is 40 °C), whereas at the bottom hole, it becomes 22 °C. In case of GPK4, there is no fluid flow along the casing zone and heat exchange for this section is lower than GPK3. Based on this, the temperature difference at the bottom hole is approximately 25 °C. Considering the wellbore heat exchange effect, colder fluid injection may have lower impact at the bottom hole temperature than expected. With increasing the flow rate, a similar trend is observed with other flow rates and a higher difference between the minimum and maximum temperature is expected. Therefore, for the lower injection flow rate, neglecting the heat exchange effect may impart huge error on the final results.

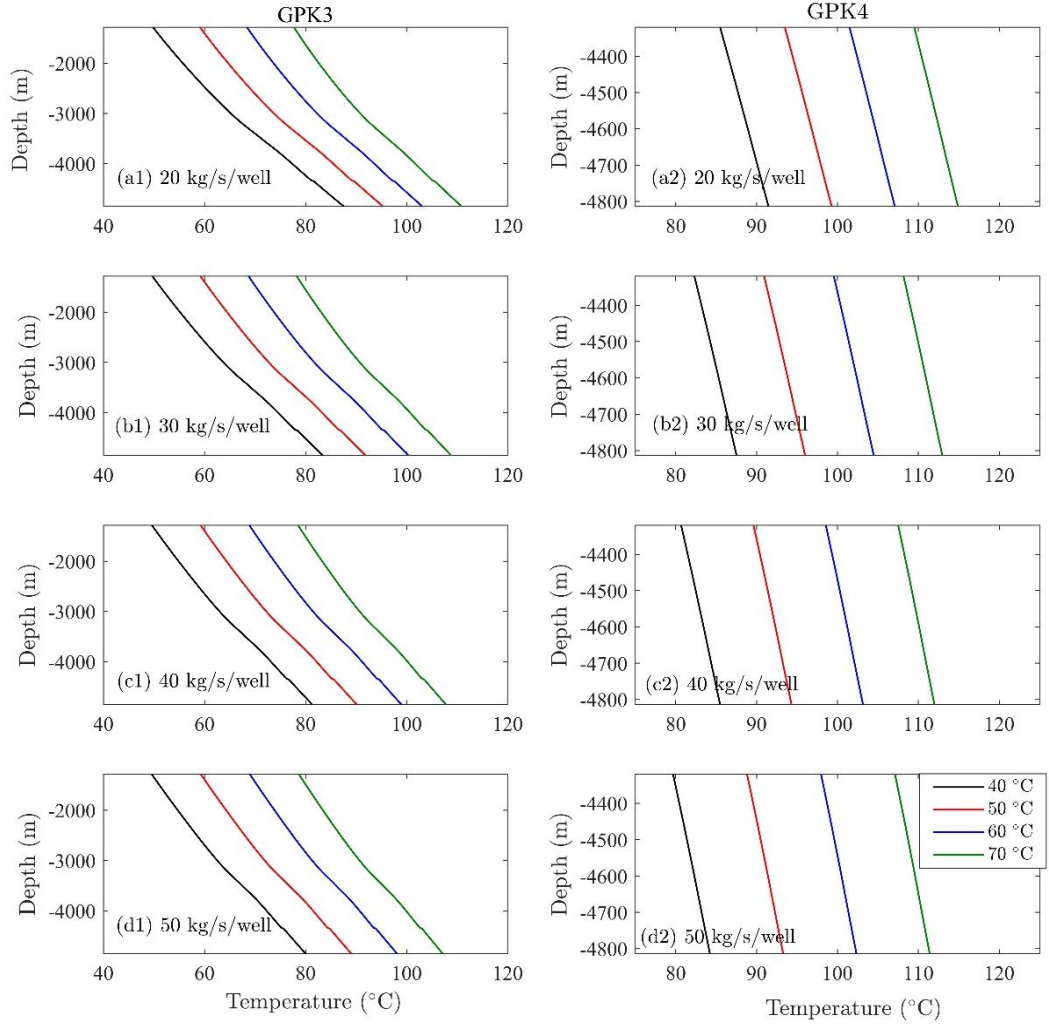

*Figure 3: Temperature profile along the injection wellbore for different injection rates. The negative sign in the depth indicates it is measured in the earth surface.*

Figure 4 shows the reservoir response towards the $CO_2$ and water injection. To obtain the temperature profile in this figure, Eq. [12] is used and considered flow rate coming from the casing zone and the open hole section of GPK2. When the fluid is $CO_2$ and the flow rate is low, the contribution from the matrix zone specially in the upper zone of the reservoir where the injection and production wells are close to each other is considerable and its initial production temperature deviates from the counterparts of the water case. By increasing the flow rates, most of the $CO_2$ comes from the bottom fault zones and show similar initial production temperature as water (see Figure 4). For all cases, at the top section of the reservoir, GPK2 and GPK3 are close to each other and heat exchange in this section reduces the production temperature sharply at early times (within 10 years of operation starting). After this period, heat exchange in this area becomes the semi-steady state and temperature reduction in the later times is mainly controlled by the deeper sections in which the production and the injection wells are present at far distance. Due to this, at later times the temperature reduction rates are smaller. However, injecting temperature differences between the maximum and minimum values for all flow rates are within 30 °C but the reservoir response for these cases eventuate to a temperature difference of less than 5 °C for water and 10 °C for $CO_2$. Therefore, the thermal breakthrough at the examined time scale is not a function of surface injection temperature. It should be noted that this temperature trend is different than thermal breakthrough. As the flow rate increases, final temperature response from the reservoir widens and it shows that lowering injection temperature at the surface will have a higher effect on the reservoir response at higher flow rates. Isosurface temperature distribution around the injection wells is depicted in Figure 5. As it is shown, the low temperature plume around GPK3 envelops the production well. Due to this, a rapid temperature reduction at early times is observed. By decreasing the injection temperature, the volume of the low temperature plume will slightly increase as expected. It is obvious from comparing Figure 5(a2 & c2) with 40 °C and Figure 5(b2 & d2) with 70 °C. Furthermore, early time (at 10 years) behavior is almost the same for different injection temperature (consider the case Figure 5(a1, b1, c1 & d1)). However, by increasing the flow rates from 20 to 50 kg/s, a significant increase in the size of the plume at 50 years is visible. Again, these plots confirm that injecting at the lower temperature at the surface has lower impact than expectation on the low temperature flow development at the surface. Since, low temperature from the plume development is a controlling factor for the thermo-elasticity and the stress field [Mahmoodpour et al. 2022b,c], this plot shows that lowering the $CO_2$ injecting temperature will not contribute much for the possibility of any significant stress development that may eventuates to a possible seismic events. Faults with higher thermal conductivity changes the area of the temperature plume and when the plume comes closer to the faults, temperature plume increases.

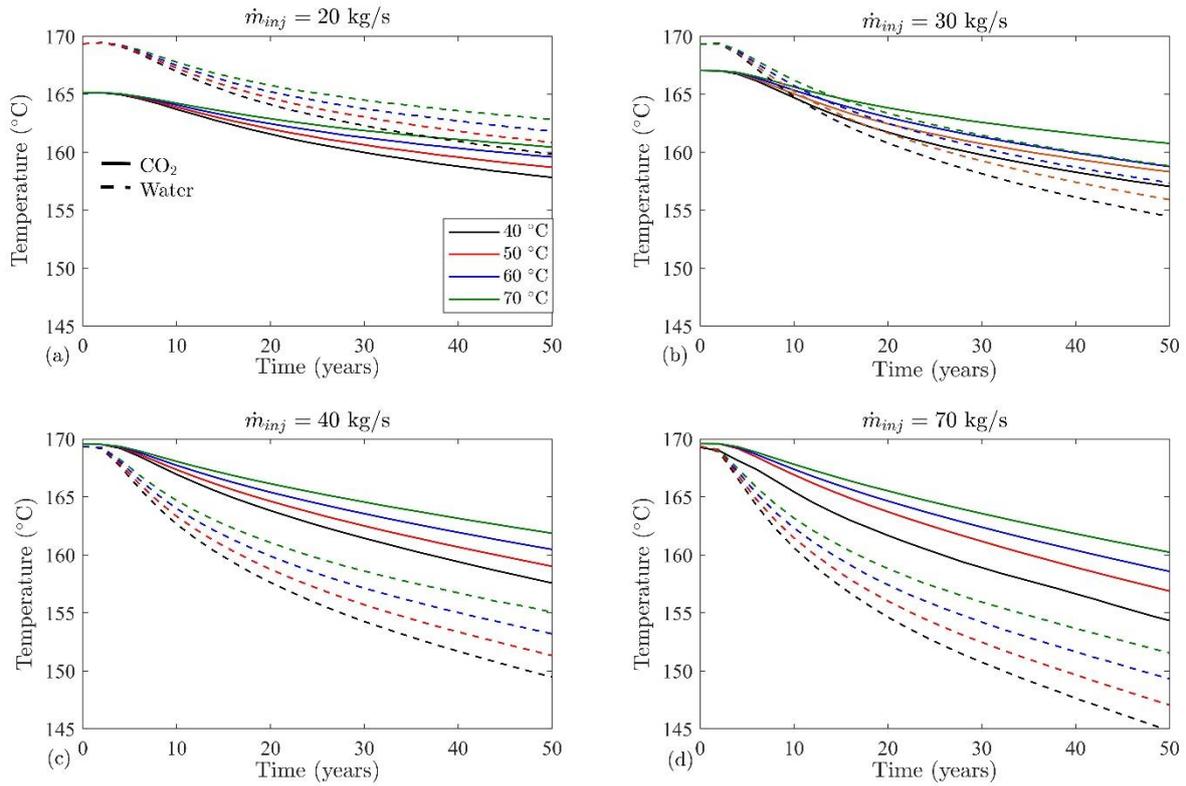

*Figure 4: Reservoir response comparison between $CO_2$ and water as working fluid for different injection rates.*

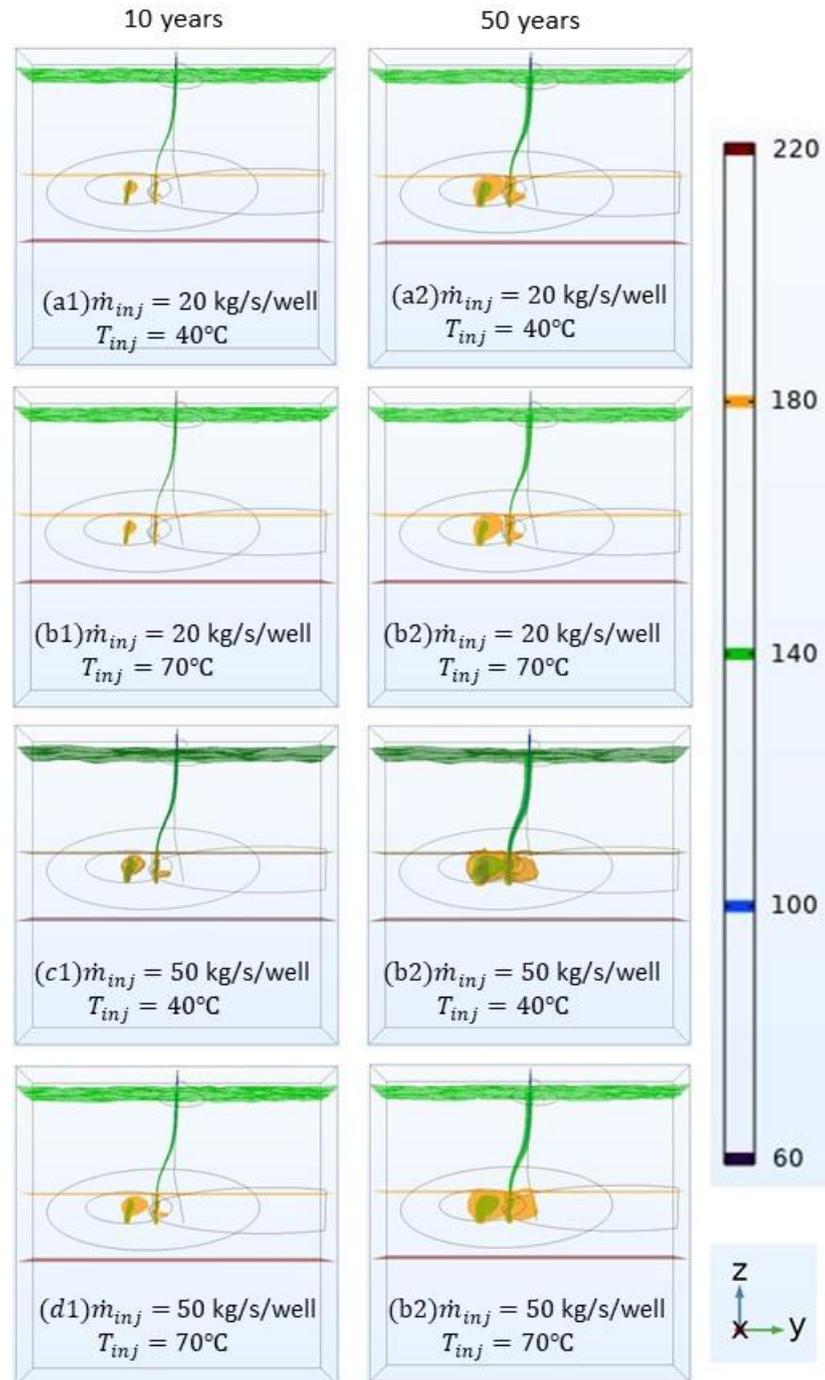

*Figure 5: Temperature isosurface development for different injection rates and injection temperatures at time 10 and 50 years.*

Figure 6 shows the temperature profile resulting from the wellbore heat exchange alongside the production well after 10 and 50 years for both the working fluids. For all flow rates and injection temperature, the temperature reduction for the $CO_2$ is higher due to its lower specific heat capacity. It should be noted that the velocity of $CO_2$ inside the wellbore is higher than water but due to the lower specific heat capacity the heat exchange happens faster. For 20 kg/s injection rate and 70 °C surface injection temperature (see Figure 6(a1)), the total temperature reduction resulting from the heat exchange along the wellbore after 10 years is almost 38 °C for $CO_2$ and 21 °C for water whereas for 50 kg/s injection rate and 70 °C surface injection temperature (see Figure 6(d1)), it is approximately 36 °C and 17 °C, respectively. This temperature drop shows that water is more sensitive to the flow rate. Same as the injection wells, by increasing the flow rate, the time for heat exchange decreases and a lower temperature reduction is observed in the production well. Since, flow rate at the production well is higher than injection wells, therefore wellbore effect on the production well is smaller. As time progresses, the difference between the bottom hole temperature and consequently temperature profiles alongside the production well increases for different injection temperatures. However, the wellbore effect on the temperature reduction is almost same after 50 years of operation and the observed values are 37 °C and 20 °C for $CO_2$ and water, respectively with the flow rate of 20 kg/s/well and the surface injection temperature as 70 °C . For the 50 kg/s/well injection rate and surface injection temperature of 70 °C, these values become 35 °C and 16 °C for $CO_2$ and water, respectively.

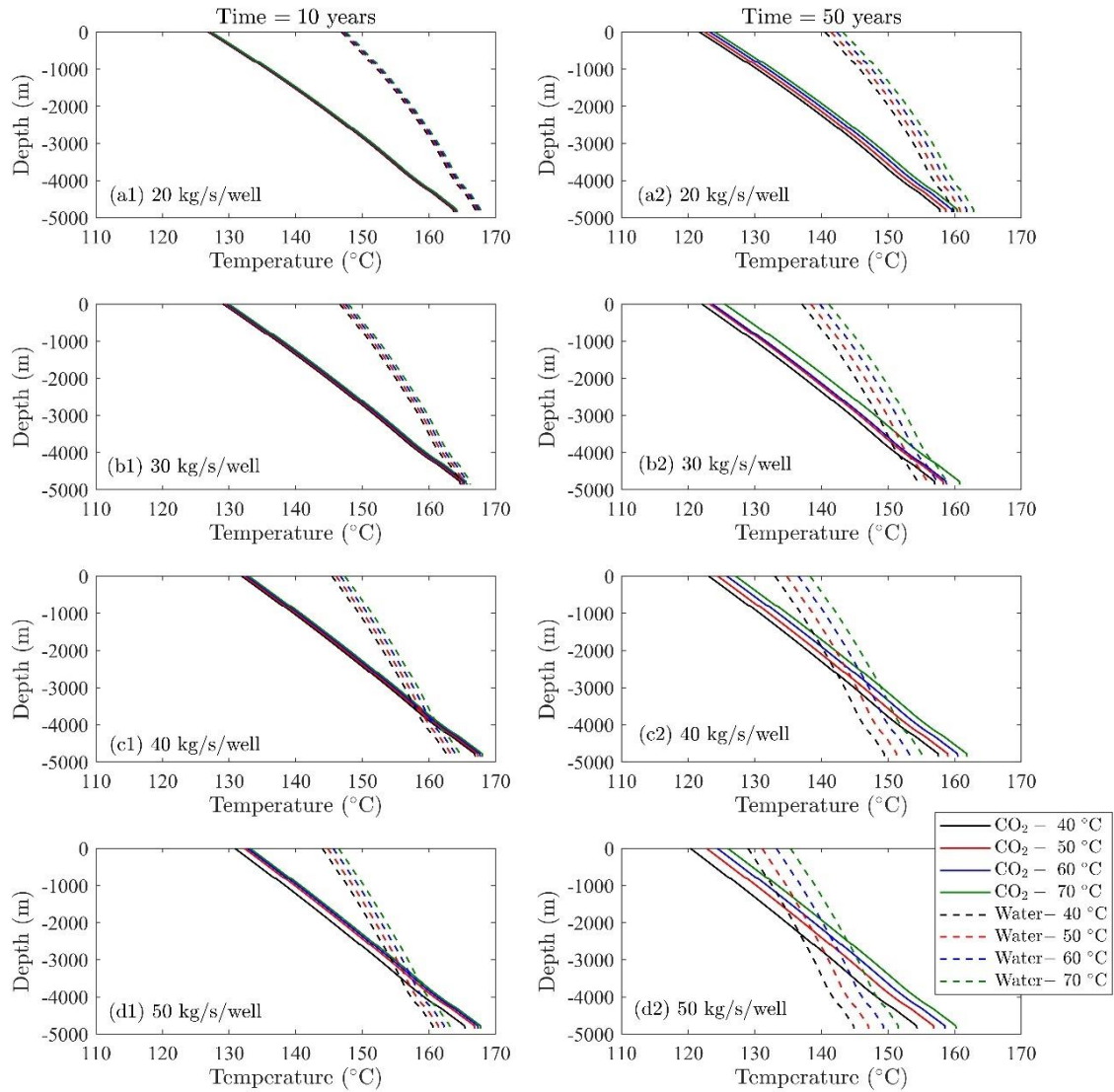

*Figure 6: Production wellbore temperature profile only due to the wellbore response for the different injection rates per well.*

Figure 7 shows the temperature alongside the production well just from the reservoir response. Figure 7(a1) shows that in the region slightly above the bottom hole of the wellbore, there is a peak minimum value in the temperature which indicates the impact of the bottom faults on the temperature reduction. In the counterpart plots for $CO_2$, Figure 7(a2), this minimum value is negligible due to lower specific heat capacity of $CO_2$. With the flow rate of 20 kg/s/well, $CO_2$ does not have the capacity to decrease the temperature of the faulted zone in the vicinity to the production wellbore. For the openhole zone (well trajectory from ~ 0 – 1500 m as shown in Figure 7), distance between the injection and the production wellbore is high enough and therefore, there is negligible impact on the production wellbore temperature in this zone. This results are much obvious for the case of $CO_2$ where after 50 years of operation, temperature is almost close to the initial reservoir temperature. However, in the upper section, where the injection and production wellbores are close to each other, the coolooing effect near the leakage zone of injection wellbore, impacts the temperature of the production wellbore. Obviously, with the lower injection temperature, this impact is higher. In case of water, near the bottom faulted zones, temperature in the production wellbore becomes higher than the initial temperature due to the higher flow rate in the faulted zone and migration of some hot water from the lower regions to this zone (see Figure 7(a1-d1)). Due to the lower specific heat capacity of $CO_2$, this region is not detectable (see Figure 7(a2-d2)). Furthermore, in the upper zone of the well (well trajectory from ~ 3000 – 3500 m as shown in Figure 7), a peak of maximum temperature happens between the two upper faulted zones due to the lower fluid velocity in this region. By increasing the flow rate, the impact of the bottom faulted zones on the production wellbore temperature increases from 10 ℃ for 20 kg/s/well case to 45 ℃ for 50 kg/s/well case with water as the working fluid. However, this impact is renounced for the case of $CO_2$ where this temperature drop occurs from 1 ℃ for 20 kg/s/well case to 26 ℃ for 50 kg/s/well case. For the 20 kg/s/well case, the differences in the temperature of the bottom faulted zone is narrow (~2 ℃) for different injection temperature of water while the 50 kg/s/well case, this difference becomes ~11℃. A similar behavior is observed for the $CO_2$ cases where, ~2 ℃ and ~10 ℃ temperature drops are estimated for 20 kg/s/well and 50 kg/s/well case, respectively. In the top leakage zone, for all flow rates and working fluids, the temperature reduction alongside the production wellbore is significant. For example, for the water as the working fluid, the cases with 20 kg/s/well flow rate, the difference between the initial temperature and the temperature in the leakage zone is ~31 ℃ and for the case with 50 kg/s/well, this temperature difference is ~41 ℃. For $CO_2$ as the working fluid, this temperature difference is ~21 ℃ and ~40 ℃ for the flow rates fo 20 kg/s/well and 50 kg/s/well, respectively. This indicates that cooling effect in the leakage zone is more sensitive to the flow rate for $CO_2$ cases.

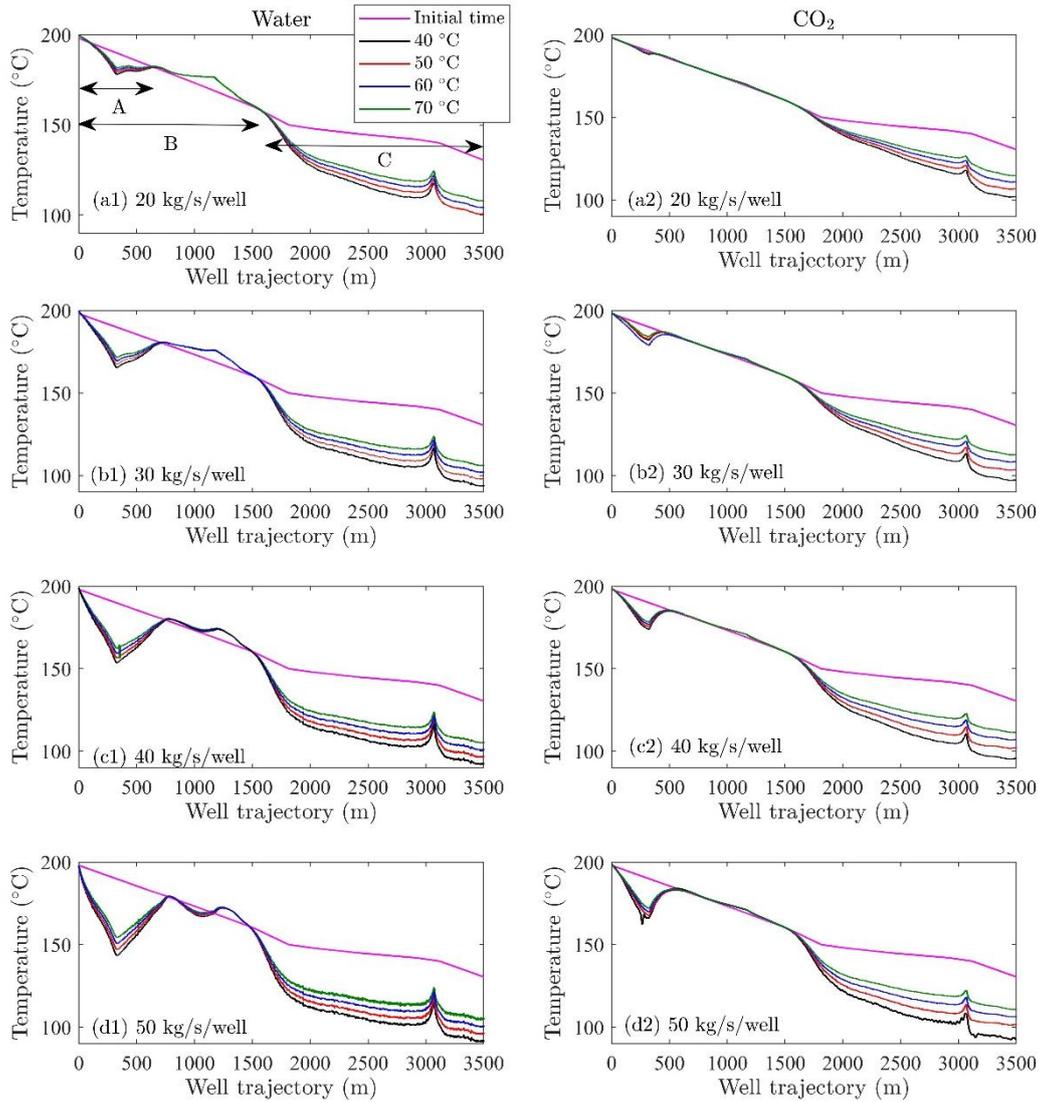

*Figure 7: Temperature profile along the production well for different injection flow rates per well. In (a1), three section of the well trajectory are categorized: A – bottom faulted zone, B – open hole section and C – top leakage zone. The x-axis indicates the well trajectory measured from the bottom of the production well.*

For the previous analysis in this study, $\Phi = 0.35$ is used. To do a sensitivity analysis on the total aspect of the process, analysis is repeated for $\Phi = 0.3$ and $\Phi = 0.4$ with two different surface injection temperatures (40 and 70 ℃) by keeping the same injection flow rates (30 kg/s). Figure 8(a1) shows that by increasing the $\Phi$ value, the impact of the wellbore heat exchange increases and the fluid reach the bottom with higher temperature. When $\Phi = 0.3$, $CO_2$ reaches the start of the cased zone with a temperature of 42 ℃ and the bottom hole with the temperature 59 ℃. Therefore, it experiences 17 ℃ increase in temperature. However, when $\Phi = 0.4$, it reached to the start of the cased zone at 57 ℃ and it leaves the bottom hole at a temperature of 106 ℃. Therefore, it produces 49 ℃ change in temperature. Similar increasing trend happen when the injection temperature is 70 ℃ (this temperature increase is 14 and 45 ℃ respectively, see Figure 7(a2)). Figure 7(b1 & b2) shows a similar trend for GPK4. Figure 7(c1 & c2) shows the production temperature as a function of reservoir response only and it indicates that by increasing the $\Phi$ value, the production temperature will increase. This is due to the fact that injecting $CO_2$ achieves higher temperature in the reservoir and cooling effect resulting from GPK3 around GPK2 is smaller. But this temperature change is marginal in compared to wellbore effect. For the production well, temperature profile increasing the $\Phi$ value, increases the heat loss alongside the wellbore. With $\Phi = 0.4$, fluid entering the wellbore with higher temperature (as explained in Figure 7 (c1 & c2)) and leaving with lower temperature (see Figure 7(d1 & d2)). In comparison to the injection wells, the behavior of temperature profile alongside the production well remain almost constant with time (see Figure 7(d1 & e1) and Figure 7(d2 & e2)) and the surface injection temperature (see Figure 7(d1 & d2) and Figure 7(e1 & e2)), but it is a strong function of

$\Phi$. Figure 7(f1) shows the temperature alongside the production well resulting from the reservoir response. As it is depicted in Figure 7(c1) that by increasing $\Phi$ values, the reservoir temperature will be higher because the injected $CO_2$ will reach with a higher temperature to the reservoir. Comparison of Figure 7(f1) and 7(f2) reveals that by increasing the injection temperature, the dependency of the temperature profile along the production well resulting from the reservoir decreases.

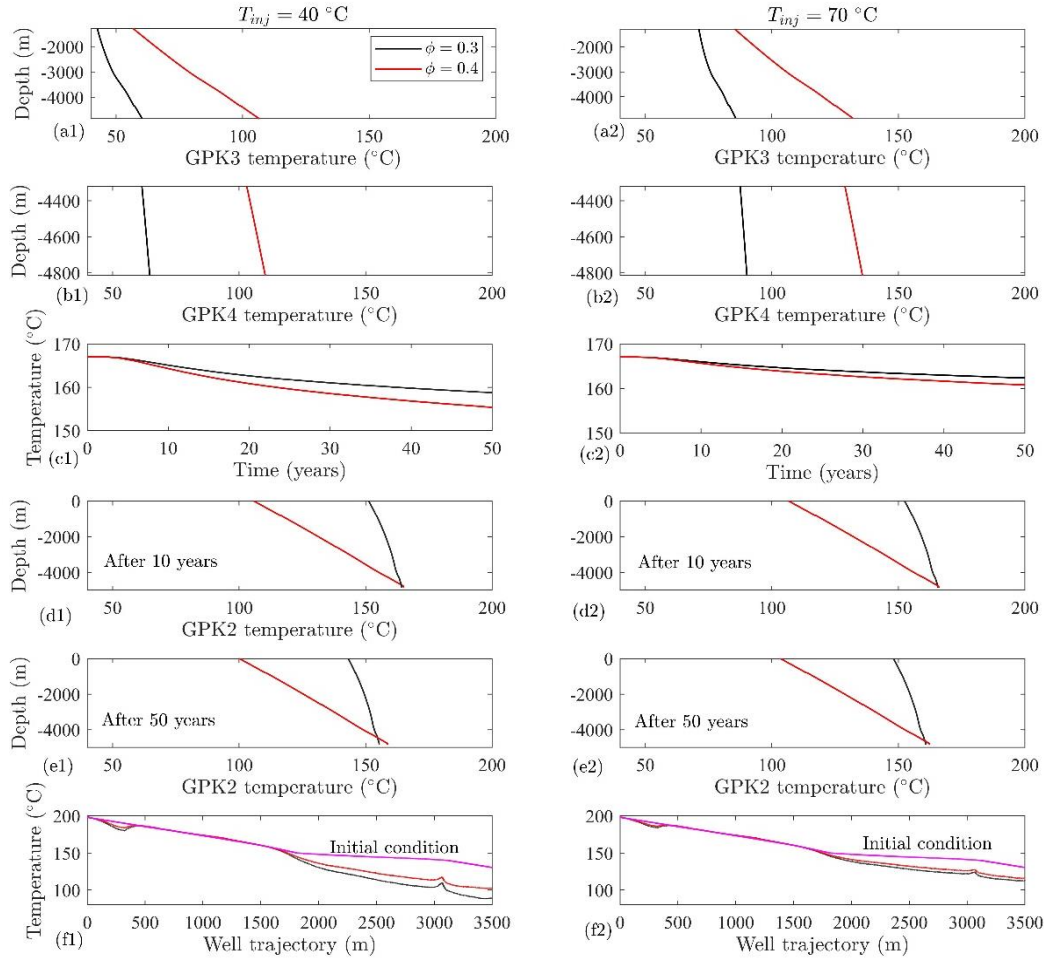

*Figure 8: Sensitivity on the wellbore heat exchange effect through the two values of $\Phi = 0.3$ and 0.4 considering $CO_2$ as the working fluid. Here flow rate is assumed as 30 kg/s for both the cases. (a1) and (a2) shows the temperature profile for GPK3 whereas (b1) and (b2) shows for GPK4. (c1) and (c2) displays the reservoir response only, (d1) and (d2) shows the wellbore heat exchange effect after 100 years whereas (e1) and (e2) shows this effect after 300 years, (f1) and (f2) represents the temperature profile alongside the production wellbore measured from the bottom hole section.*

**Conclusions**

A hydro-thermal reservoir scale numerical simulation is developed for the Soultz-sous-Forêts by using supercritical $CO_2$ as the working fluid for heat exchange. It is observed that cold $CO_2$ re-injection may eventuate to a small production temperature reduction at least for Soultz-sous-Forêts case. Therefore, injection temperature may not have great impact on the bottom hole temperature. Considering the wellbore heat exchange effect, it is essential to get accurate estimation of the heat extraction potential for geothermal reservoir systems. The low temperature plume development surrounding the injection well is the primary contributor to the temperature reduction in the production wellbore. In comparison to the water, $CO_2$ shows lower temperature reduction in the faulted and the leakage zone for 50 years of operation, which makes it a suitable working fluid. This study suggests that sufficient spacing between injection and production wellbores for new field developments. Temperature changes resulting from the wellbore may have multiple times higher impact on the production temperature in comparison to the reservoir response. Wellbore response is a strong function of lumped parameter for kinetic energy and Joule-Thomson coefficient which accounts for the casing properties, cement properties and their thicknesses. Therefore, the lumped parameter should be calculated based on the specific properties of each wellbore and the working fluid or through the production temperature data matching.

**Author Contributions:** Conceptualization, M. Singh, S. Mahmoodpour and I. Sass; methodology, M. Singh and S. Mahmoodpour.; software, M. Singh and S. Mahmoodpour; validation, M. Singh and S. Mahmoodpour; writing—original draft preparation, M. Singh and S. Mahmoodpour; writing—review and editing, M.R. Soltanian, R. Ershadnia; visualization, M. Singh and S. Mahmoodpour; supervision, M.R. Soltanian and I. Sass; project administration, not applicable; funding acquisition, not applicable. All authors have read and agreed to the published version of the manuscript.

**Funding:** Authors have received financial and technical support from the Group of Geothermal Science and Technology, Institute of Applied Geosciences, Technische Universität Darmstadt.

**Conflicts of Interest:** The authors declare no conflict of interest.

**References**
Adams, B.M., Kuehn, T.H., Bielicki, J.M., Randolph, J.B. and Saar, M.O. On the importance of the thermosiphon effect in CPG ($CO_2$ plume geothermal) power systems. Energy, 2014, 69, pp.409-418.

Baillieux, O.; Schill, E.; Edel, J.-B.; Mauri, G. Localization of temperature anomalies in the Upper Rhine Graben: insights from geophysics and neotectonics activity. Int. Geol. Rev. 2013, 55, 1744-1762.

Bai, B.; He, Y.; Li, X.; Numerical study on the heat transfer characteristics between supercritical carbon dioxide and granite fracture wall. Geothermics, 2018, 75:40–47.

Blumenthal, M.; Kuhn, M.; Pape, H.; Rath, V.; Clauser, C. Hydraulic model of the deep reservoir quantifying the multi-well tracer test. Proceedings of the EHDRA scientific conference, Soultz-sous-Forêts, France, 28-29 June 2007, (2007).

Brown, D. A hot dry rock geothermal energy concept utilizing supercritical $CO_2$ instead of water. In 25th workshop on geothermal reservoir engineering, 2000, Stanford University, Stanford. SGP-TR-165.

Buscheck, T.A., Bielicki, J.M., Edmunds, T.A., Hao, Y., Sun, Y., Randolph, J.B. and Saar, M.O. Multifluid geo-energy systems: Using geologic $CO_2$ storage for geothermal energy production and grid-scale energy storage in sedimentary basins. Geosphere, 2016, 12(3), pp.678-696.

Chen, Y.; Ma, G.; Wang, H.; Li, T.; Wang, Y. Application of carbon dioxide as working fluid in geothermal development considering a complex fractured system. Energ Convers Manag, 2019, 180:1055–1067.

Cocherie, A.; Guerrot, C.; Fanning, C.M.; Genter, A. Datation U–Pb Des Deux Faciès Du Granite de Soultz (Fossé Rhénan, France). Comptes Rendus Geosci., 2004, 336, 775–787.

COMSOL Multiphysics® v. 5.5. www.comsol.com. COMSOL AB, Stockholm, Sweden.

Dezayes, C.; Genter, A.; Valley, B. Structure of the low permeable naturally fractured geothermal reservoir at Soultz. C. R. Geoscience, 2010, 342, 517–530. https://doi.org/10.1016/j.crte.2009.10.002.

Duringer, P.; Aichholzer, C.; Orciani, S.; Genter, A. The Complete Lithostratigraphic Section of the Geothermal Wells in Rittershoffen (Upper Rhine Graben, Eastern France): A Key for Future Geothermal Wells. BSGF Earth Sci. Bull. 2019, 190, 13.

Egert, R.; Maziar, G.K.; Held, S.; Kohl, T. Implications on large-scale flow of the fractured EGS reservoir Soultz inferred from hydraulic data and tracer experiments. Geothermics, 2020, 84, 101749.

Ezekiel, J., Kumbhat, D., Ebigbo, A., Adams, B.M. and Saar, M.O. Sensitivity of reservoir and operational parameters on the energy extraction performance of combined $CO_2$-EGR–CPG systems. Energies, 2021, 14(19), p.6122.

Ezekiel, J., Adams, B.M., Saar, M.O. and Ebigbo, A. Numerical analysis and optimization of the performance of $CO_2$-Plume Geothermal (CPG) production wells and implications for electric power generation. Geothermics, 2022, 98, 102270.

Gentier, S.; Rachez, X.; Ngoc, T. D. T.; Peter-Borie, M.; Souque, C. 3D flow modelling of the medium-term circulation test performed in the deep geothermal site of Soultz-sous-Forêts (France). 2010.

Gessner, K.; Kuhn, M.; Rath, V.; Kosack, C.; Blumenthal, M.; Clauser, C. Coupled process models as a tool for analysing hydrothermal systems. Surveys in Geophysics, 2009, 30.

Hasan, A.R.; Kabir, C.S.; Sarica, C. Fluid flow and heat transfer in wellbores. Society of Petroleum Engineers, 2002, Richardson.

Hasan, A.R.; Kabir, C.S.; Wang, X. A robust steady-state model for flowing fluid temperature in complex wells. SPE Prod. Operat., May 2009, 269-276.

Hou, L., Yu, Z., Luo, X., Wu, S. Self-sealing of caprocks during $CO_2$ geological sequestration. Energy, 2022, 252, 124064.

IPCC, 2022: Climate Change 2022: Impacts, Adaptation, and Vulnerability. Contribution of Working Group II to the Sixth Assessment Report of the Intergovernmental Panel on Climate Change [H.-O. Pörtner, D.C. Roberts, M. Tignor, E.S. Poloczanska, K. Mintenbeck, A. Alegría, M. Craig, S. Langsdorf, S. Löschke, V. Möller, A. Okem, B. Rama (eds.)]. Cambridge University Press. In Press.

Liu, L.; Suto, Y.; Bignall, G.; Yamasaki, N.; Hashida, T. $CO_2$ injection to granite and sandstone in experimental rock/hot water systems. Energy Convers Manag, 2003, 44(9):1399–1410.

Liu, Y.; Wang, G.; Yue, G.; Lu, C.; Zhu, X.; Impact of $CO_2$ injection rate on heat extraction at the HDR geothermal field of Zhacanggou China. Environ Earth Sci, 2017, 76(6):1–11.

Luo, F., Xu, R.N. and Jiang, P.X. Numerical study of the influence of injection/production well perforation location on $CO_2$-EGS system. Energy Procedia, 2013, 37, pp.6636-6643.

Luo, F.; Xu, R.N.; Jiang, P.X. Numerical investigation of fluid flow and heat transfer in a doublet enhanced geothermal system with $CO_2$ as the working fluid ($CO_2$ -EGS). Energy, 2014, 64:307–322.

Magnenet, V.; Fond, C.; Genter, A.; Schmittbuhl, J. Two-dimensional THM modelling of the large scale natural hydrothermal circulation at Soultz-sous-Forêts. Geothermal Energy, 2014, https://link.springer.com/article/10.1186/s40517-014-0017-x.

Mahmoodpour, S.; Rostami, B.; Soltanian, M.R.; Amooie, A. Convective dissolution of carbon dioxide in deep saline aquifers: Insights from engineering a high-pressure porous cell. Phys. Rev. Appl., 2019a, 12, 034016.

Mahmoodpour, S.; Rostami, B.; Soltanian, M.R.; Amooie, A. Effect of brine composition on the onset of convection during $CO_2$ dissolution in brine. Computers and Geosciences, 2019b, 124, 1-13.

Mahmoodpour, S.; Singh, M.; Turan, A.; Bär, K.; Sass, I. Hydro-thermal modeling for geothermal energy extraction from Soultz-sous-Forêts, France, Geosciences, 2021, 11(11), 464.

Mahmoodpour, S., Singh, M., Turan, A., Bär, K., Sass, I., Simulation and global sensitivity analysis of the thermo-hydraulic-mechanical processes in a fractured geothermal reservoir, Energy, 2022a, 247, 123511.

Mahmoodpour, S., Singh, M., Bär, K., Sass, I., Thermo-hydro-mechanical modeling of an Enhanced geothermal system in a fractured reservoir using $CO_2$ as heat transmission fluid- A sensitivity investigation, Energy, 2022b, 124266.

Mahmoodpour, S., Singh, M., Mahyapour, R., Tangirala, S.K., Bär, K. and Sass, I., 2022c. Direct numerical simulation of thermo-hydro-mechanical processes at Soultz-sous-Forêts. arXiv preprint arXiv:2206.01830.

Moradi, B.; Awang, M.B. Heat transfer in the formation. Res J Appl Sci Eng Technol, 2013, 6(21):3927–3932.

Moradi, B.; Ayoub, M.; Bataee, M.; Mohammadian, E. Calculation of temperature profile in injection wells. J Petrol. Explor. Prod. Tech. 2020, 10, 687-697.

Pruess, K. Enhanced geothermal systems (EGS) using $CO_2$ as working fluid - a novel approach for generating renewable energy with simultaneous sequestration of carbon. Geothermics, 2006, 35:351–367.

Pruess, K. On production behavior of enhanced geothermal systems with $CO_2$ as working fluid. Energy Convers Manag, 2008, 49(6):1446–1454.

Rolin, P.; Hehn, R.; Dalmais, E.; Genter, A. D3.3 Hydrothermal model matching colder reinjection design, WP3: Upscaling of thermal power production and optimized operation of EGS plants. 2018, H2020 Grant Agreement No-792037.

Sanjuan, B.; Pinault, J.L.; Rose, P.; Gerard, A.; Brach, M.; Braibant, G.; Crouzet, C.; Foucher, J.C.; Gautier, A. Geochemical fluid char-acteristics and main achievements about traces tests at Soultz-sous-Forêts (France). EHDRA Scientific Conference 2006, Soultz-sous-Forêts, France; 2006.

Sausse, J.; Dezayes, C.; Dorbath, L.; Genter, A.; Place, J. 3D model of fracture zones at Soultz-sous-Forêts based on geological data, image logs, induced microseismicity and vertical seismic profiles. C. R. Geoscience, 2010, 342, 531-545, https://doi.org/10.1016/j.crte.2010.01.011.

Schill, E.; Genter, A.; Cuenot, N.; Kohl, T. Hydraulic performance history at the Sultz EGS reservoirs from stimulation and long-term circulation tests. Geothermics, 2017, 70, 110-124.

Shaik, A.R.; Rahman, S.S.; Tran, N.H.; Tran, T. Numerical simulation of fluid -rock coupling heat transfer in naturally fractured geothermal system. Applied Thermal Engineering 2011, 31, 1600-1606.

Singh, M.; Chaudhuri, A.; Stauffer, P.H.; Pawar, R.J. Simulation of gravitational instability and thermo-solutal convection during the dissolution of $CO_2$ in deep storage reservoirs Wat. Resour. Resear., 2020a, 56, e2019WR026126.

Singh, M.; Chaudhuri, A.; Soltanian, M.R.; Stauffer, P.H. Coupled multiphase flow and transport simulation to model $CO_2$ dissolution and local capillary trapping in permeability and capillary heterogeneous reservoir. Int. J. Green. Gas. Cont., 2021, 108, 103329.

Singh, M.; Tangirala, S.K.; Chaudhuri, A. Potential of $CO_2$ based geothermal energy extraction from hot sedimentary and dry rock reservoirs, and enabling carbon geo-sequestration. Geom. Geop. Geo-Energy Geo-resour., 2020b, 6, 16.

Span, R. and Wagner, W.; A New Equation of State for Carbon Dioxide Covering the Fluid Region from the Triple-Point Temperature to 1100 K at Pressures up to 800 MPa. J. Phys. Chem. Ref. Data, 25(6):1509-1596, 1996.

Vallier, B; Magnenet, V.; Schmittbuhl, J.; Fond, C. Large scale hydro-thermal circulation in the deep geothermal reservoir of Soultz-sous-Forêts. Geothermics, 2020, 78, 154-169.

Wang, Y., Li, T., Chen, Y., Ma, G. Numerical analysis of heat mining and geological carbon sequestration in supercritical $CO_2$ circulating enhanced geothermal systems inlayed with complex discrete fracture networks. Energy, 2019, 173, 92-108.

Zhang, L.; Luo, F.; Xu, R.; Jiang, P.; Liu, H. Heat transfer and fluid transport of supercritical $CO_2$ in enhanced geothermal system with local thermal non-equilibrium model. Energy Procedia, 2014, 63:7644–7650.